\documentclass[runningheads]{llncs}
\usepackage[T1]{fontenc}
\usepackage[nocompress]{cite}
\usepackage{xcolor}
\usepackage{comment}
\usepackage[hyphens]{url}
\usepackage{graphicx}
\usepackage{booktabs}
\usepackage[svgnames,table]{xcolor}
\usepackage{pifont}
\usepackage{dirtytalk}
\usepackage{enumitem}
\usepackage[nointegrals]{wasysym}
\usepackage{bibunits}
\usepackage{arydshln}
\usepackage{tabularx}
\usepackage{rotating}
\usepackage{amssymb}
\usepackage{pdfrender}
\usepackage{soul}
\usepackage{subfiles}
\usepackage{tikz}
\usepackage{float}
\graphicspath{{Images/}}
\usepackage{boxhandler}
\usepackage{comment}
\usepackage{tablefootnote}
\usepackage{amssymb}
\usepackage{blindtext}
\usepackage[normalem]{ulem}
\usepackage{graphicx}
\usepackage{caption}
\usepackage{array} %
\usepackage{hyperref}
\usepackage{graphicx}
\usepackage{multirow} %

\usepackage[caption=false,font=footnotesize]{subfig}
\usepackage{graphicx}

\usepackage{subfig}  %
\spnewtheorem{condition}{Condition}{\bfseries}{\itshape}

\usepackage[most]{tcolorbox} %
\usepackage{pifont}%

\begin{document}
\title{Stress Testing Social Media Authentication Ceremonies of End-to-End Encrypted Messaging with Ephemeral Fake Accounts}
\titlerunning{Stress Testing Social Media Authentication Ceremonies}

\author{Mashari Alatawi\inst{1} \and
Md Shahidur Rahaman\inst{2}\thanks{Corresponding author.} \and
Nitesh Saxena\inst{2}}
\authorrunning{M. Alatawi et al.}
\institute{King Saud University, Riyadh, Saudi Arabia\\
\email{Malatawi@ksu.edu.sa} \and
Texas A\&M University, College Station, TX, USA\\
\email{\{mdshahidur\_rahaman,nsaxena\}@tamu.edu}}

\newcommand{\attackname}{SocFakeID}

\maketitle              %
\begin{abstract}
The authentication ceremony in end-to-end encrypted messaging verifies the binding between a user's identity and their public key, preserving security even against an adversarial, compromised, or coerced key server. To improve usability, recent proposals replace opaque fingerprint comparison with social media-based verification, as in Vaziripour et al.\ (CHI'19) and Linker and Basin (USENIX Sec'24). We show that this design decision introduces a structural vulnerability: by placing identity selection under service control, social authentication transfers the trust anchor from a cryptographic primitive to the very service it is meant to constrain. We formalize the single condition a ceremony must satisfy to resist an adversarial service, and introduce \attackname, an attack that violates it by construction. A compromised or coerced key server binds attacker-controlled social media accounts to forged public keys and delivers them as the identity evidence shown during the ceremony, requiring no credential compromise, no off-platform lure, and operating entirely within the trusted application interface. We evaluate \attackname\ in a controlled study of 30 participants drawn from a technically skilled, security-aware cohort, the group least likely to fail, and observe a 93.33\% attack success rate against 2\% for clearly mismatched profiles. Even participants who inspected the linked accounts failed at high rates, indicating that the failure stems from the ceremony's interface constraints rather than from user inattention. Our findings show that social authentication ceremonies cannot provide meaningful security guarantees when the service controls identity presentation and the interface surfaces only forgeable cues.

\end{abstract}
\section{Introduction}
End-to-end encryption defends communication even against a compromised, malicious, or coerced key server, but only if users verify that received public keys belong to their intended peers~\cite{impersonation2024attack,zhang2009feasibility}. This verification, the authentication ceremony, is typically a manual fingerprint comparison that users routinely skip or botch~\cite{alatawi2023sok}. To make it usable, researchers bound key verification to social media identities~\cite{vaziripour2019don,keybase,linker2024soap}. Keybase did so through user-controlled public posts any third party can verify outside the messaging service. Vaziripour et al.~\cite{vaziripour2019don} and Linker and Basin~\cite{linker2024soap} instead built social authentication into E2EE apps, trading fingerprint comparison for a new dependency: the integrity of the identity cues the service surfaces during the ceremony. Vaziripour et al.~\cite{vaziripour2019don} found the approach convenient but noted users barely inspected linked accounts, treating this as a platform problem of fake-account detection. That diagnosis is wrong. The weakness is architectural: the service, not the platform, decides which identities appear during the ceremony, and even flawless platform-side detection is moot when the adversarial service is itself the channel delivering those identities.

\textbf{The Core Problem.} Social authentication does not solve the ceremony's verification problem; it relocates it to worse ground. Fingerprint comparison asks a question with a deterministic answer: do these bytes match? Social authentication asks an ill-defined one: do these accounts genuinely belong to my peer? The user cannot answer it without prior out-of-band knowledge of the peer's social presence, knowledge absent in precisely the settings E2EE authentication exists to protect: new professional contacts, journalists and their sources, pseudonymous and cross-platform exchanges. Against an adversarial service, the very threat E2EE authentication is built to counter, the ceremony's security collapses to trusting whatever identities the service elects to show. We formalize this as a structural condition: a social authentication ceremony resists an adversarial service only if the service delivers identities faithfully, or the user already holds an independent reference to the peer's identity.

We show that social authentication ceremonies carry a systematic identity-binding weakness wherever the service can manipulate the mapping from a peer to the identities shown for verification, and we instantiate it as \attackname. A malicious key directory binds attacker-controlled accounts to the intended peer mid-ceremony, leading users to accept attacker keys with no credential theft and no off-platform lure. Unlike phishing, \attackname\ exploits no implementation flaw but the social authentication architecture itself, using short-lived fake accounts that evade standard platform detection~\cite{ephemeral2020facebook,ephemeral2023mindy,ephemeral2023so}. A successful attack hands the adversary the full E2EE session, intercepting and manipulating messages, voice, and video (Section~\ref{study4_threat_model}).

\begin{tcolorbox}[colback=gray!10!white, colframe=black!50!white,
  boxsep=2pt, left=2pt, right=2pt, top=2pt, bottom=2pt]
\attackname\ exploits a structural weakness in social authentication ceremonies whose interface surfaces only forgeable identity cues (e.g., Vaziripour et al.\ (CHI'19) and Linker-Basin (USENIX Sec'24)). \textbf{It works entirely within the standard E2EE threat model, which assumes the service provider or key server can be compromised, malicious, or under coercion.} The social media provider is still assumed honest. \textbf{The attack requires no compromise of user credentials at any provider.} Platform-side fake account detection does not mitigate it, as we detail in Section~\ref{sec:attack}.
\end{tcolorbox}

We evaluate \attackname\ in a controlled study. When the attacker substitutes fake accounts during the ceremony, users fail to detect it with high probability. Our participant pool consisted of technically skilled, security-aware individuals with high social media fluency, making our findings a conservative lower bound: failure rates in broader populations are likely at least as severe. The vulnerability is therefore not a product of user inattentiveness, but of the constraints imposed by service-controlled identity presentation.

\noindent\textbf{Our Contributions.}
\begin{itemize}[noitemsep,topsep=0pt,parsep=0pt,partopsep=0pt, leftmargin=*]
\item \textbf{A structural security condition for social authentication.} We formalize the single condition (Identity Binding Integrity, Condition~\ref{cond:ibi}) a ceremony must satisfy to resist an adversarial service, and prove via Proposition~\ref{prop:violation} that \attackname\ violates it by construction even when every platform is fully honest. This locates the weakness at the identity-presentation layer, not in platform-side fake-account detection~\cite{vaziripour2019don}.

\item \textbf{Instantiation of the \attackname\ attack.} We realize the condition's violation as a concrete attack: short-lived attacker-controlled accounts that mimic the peer's public identities and are delivered as ceremony evidence entirely within the trusted interface, with no external lure or phishing. The instantiation shows that attacker effort is minimal once the service controls the presented identity set, confirming the threat is practical rather than purely theoretical.

\item \textbf{Empirical evaluation of ceremony failure.} We evaluate \attackname\ in a controlled study of a technically skilled, security-aware cohort, the group least likely to be fooled. Participants accepted attacker-controlled identities in 93.33\% of trials (95\% CI: [77.9\%, 99.2\%]) while rejecting clearly mismatched profiles (2\% acceptance). The large gap shows that low-effort substitutions pass inspection within the ceremony's interface constraints, making the observed rate a conservative lower bound on attack effectiveness.
\end{itemize}

\textbf{Artifact Availability:} All artifacts, the Android application, simulated malicious keyserver, anonymized trial dataset, and analysis scripts, are available at \href{https://sites.google.com/view/socfakeid-v2/artifacts}{https://sites.google.com/view/socfakeid-v2/artifacts} . 

\section{Background}
\subsection{End-to-End Encrypted Applications}
End-to-end encrypted applications protect private messaging, voice and video calls, and file sharing by encoding messages on the sender's device and decrypting them only on the recipient's~\cite{borisov2004off,frosch2016secure,cohn2020formal,zimmermann2011zrtp,ermoshina2016end}. Many rely on the Signal protocol~\cite{frosch2016secure}, including Signal~\cite{signal}, WhatsApp~\cite{whatsapp}, Google Messages~\cite{messagesbygoogle}, Skype~\cite{skype}, and Facebook Messenger~\cite{facebookmessenger}, while others such as Telegram~\cite{telegram} and Threema~\cite{threema} use proprietary protocols that exchange public keys through a central server. Because key exchange and message routing pass through a central server, users must verify that the keys they receive are authentic and not substituted by an attacker~\cite{herzberg2016can,yadav2022automatic}. To this end, most such applications provide dedicated authentication ceremonies that confirm the legitimacy of exchanged public keys~\cite{alatawi2023sok}.

\subsection{Social Media--Based Authentication}
\label{lb:socialmedia}

When users install an end-to-end encrypted application, it generates a public and private key pair; the public key is stored on a central key server and the private key remains on the device. Users depend on this server to obtain each other's public keys, a model known as opportunistic end-to-end encryption~\cite{langley2009opportunistic,unger2015sok,herzberg2016can}, which is secure only if the server behaves honestly. A compromised or malicious server can perform a key substitution attack by replacing real keys with attacker-controlled ones~\cite{herzberg2016can,yadav2022automatic}, enabling a man-in-the-middle attacker to intercept, modify, or impersonate communications. To defend against this, many applications require an authentication ceremony in which users confirm authenticity by scanning QR codes, comparing numeric values, or matching emojis. Prior work shows users often struggle with these tasks~\cite{shirvanian2015security,assal2015s,herzberg2016can,schroder2016signal,shirvanian2017pitfalls,vaziripour2017you,dechand2016empirical,tan2017can}.

\begin{figure}[htbp]
	\centering
	\includegraphics[width=0.35\textwidth, height=7cm]{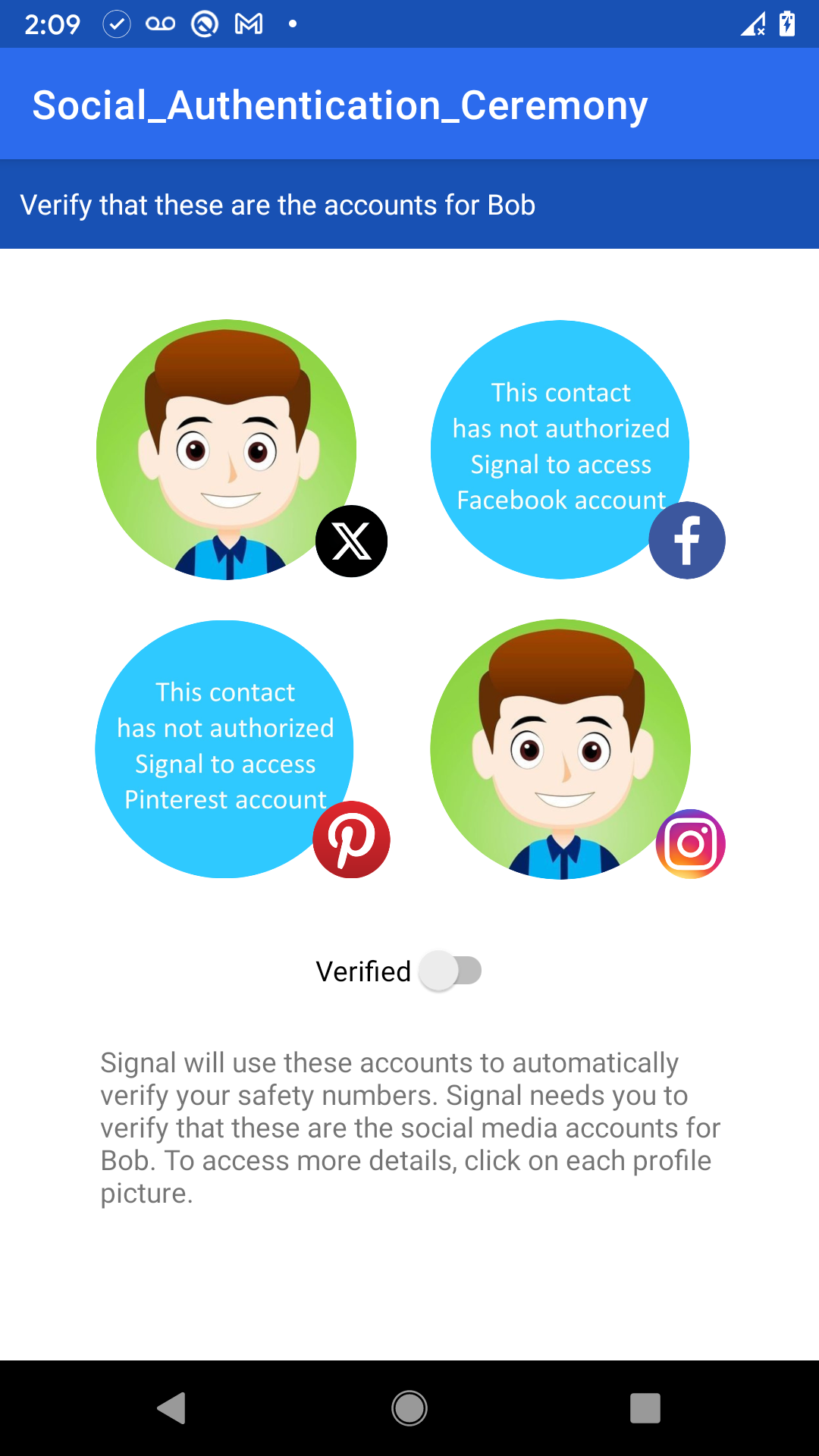}
	\caption{UI of authentication ceremony using social media networks~\cite{vaziripour2019don}}
	\label{study4pics_fig2}
\end{figure}

To improve usability, Vaziripour et al.~\cite{vaziripour2019don} introduced social media based authentication, inspired by Keybase~\cite{keybase}. Keybase binds public keys to social media accounts through \emph{user-controlled} public posts any third party can verify without a central directory, keeping the identity-binding proof under the user's control. Vaziripour et al.'s method instead presents the peer's social media profiles during the ceremony, showing a profile picture per account for the user to confirm with a toggle (Figure~\ref{study4pics_fig2}), then checks that the keys those profiles advertise match the server's. Their study focused on usability and found the approach convenient, but noted users barely inspected linked accounts, attributing this to platforms' responsibility for fake-account detection. As we show, the concern is architectural, not platform-dependent: the service, not the user, controls which identities appear during the ceremony.

\begin{figure}[!t]
\centering
\subfloat[Prover's identity and safety number shown in the web-based prototype.
\label{study4pics_fig_soap_web}]{
    \includegraphics[width=0.45\columnwidth]{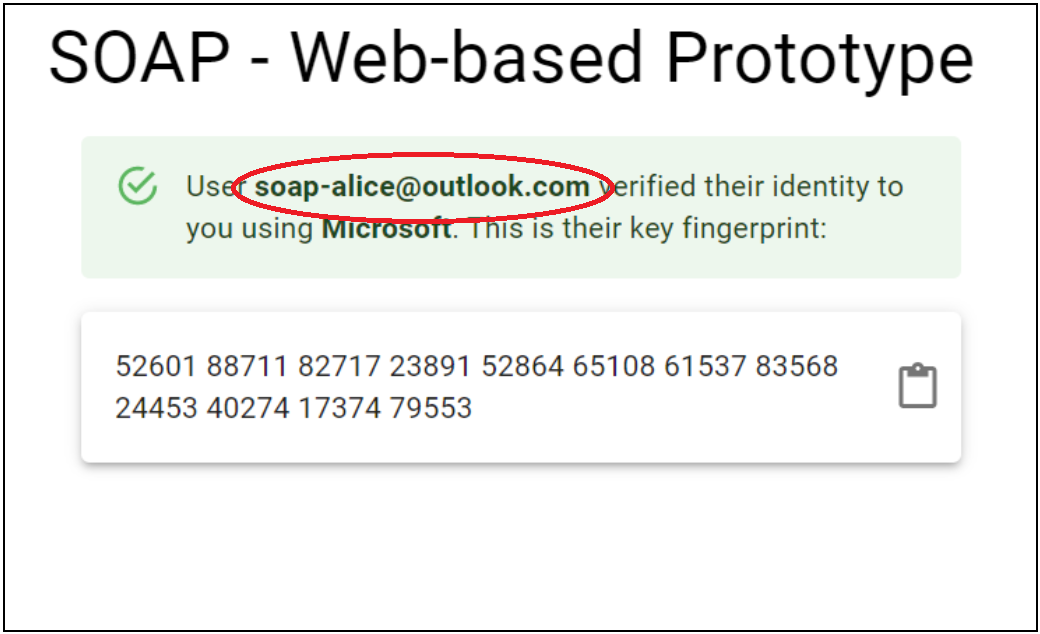}
}
\hfill
\subfloat[Signal's interface displaying prover identity during authentication.
\label{study4pics_fig_soap_signal}]{
    \includegraphics[width=0.49\columnwidth]{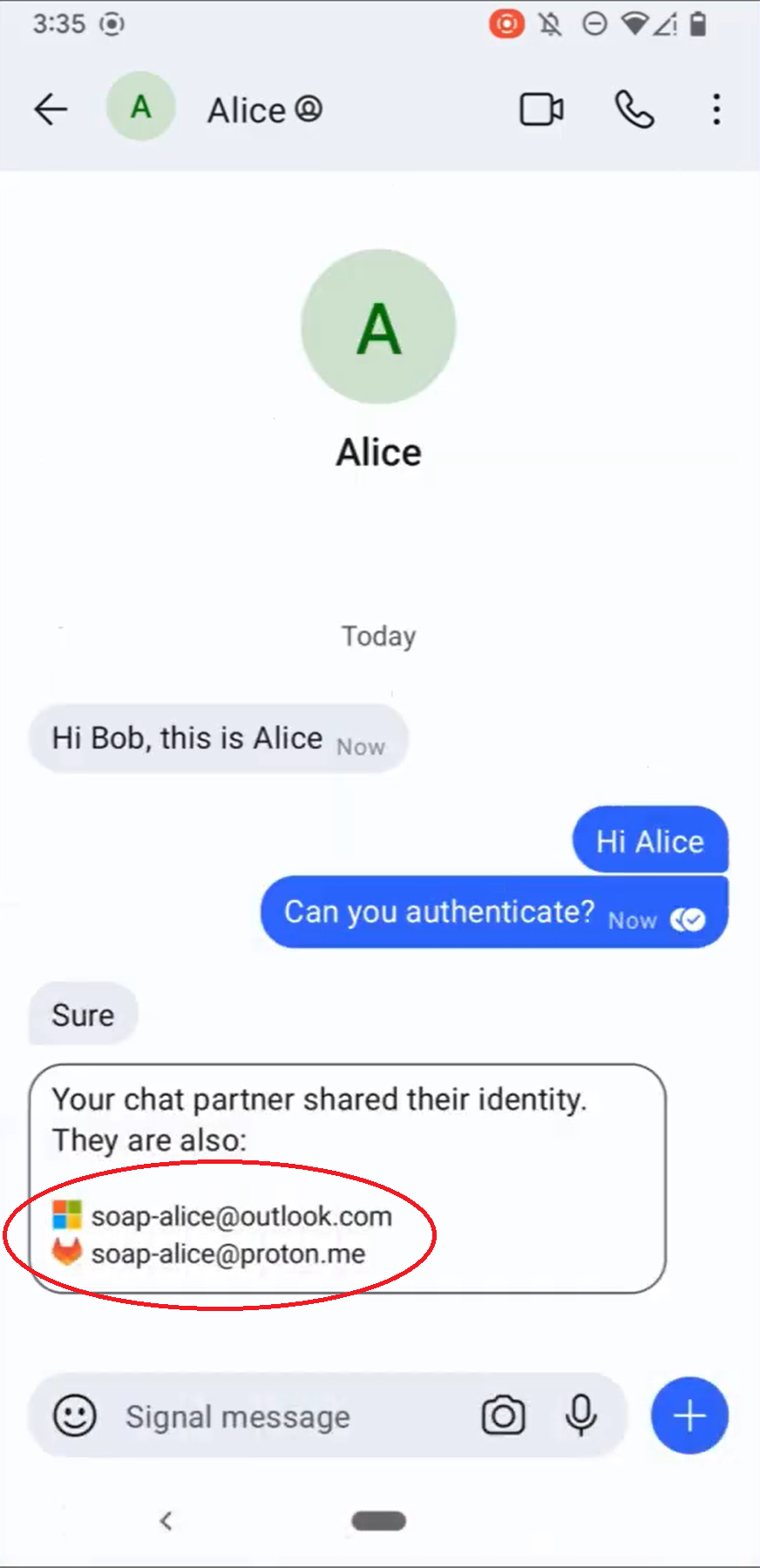}
}
\caption{UI examples of social authentication ceremonies adapted from Linker and Basin~\cite{linker2024soap}.}
\label{study4pics_fig_soap}
\end{figure}

Linker and Basin~\cite{linker2024soap} proposed SOAP, in which chat partners authenticate peers using digital identities managed by identity providers (IdPs) via OpenID Connect. They demonstrated two prototypes, web-based and Signal-integrated, that verify IdP-issued tokens and display the shared identities for comparison (Figures~\ref{study4pics_fig_soap_web} and~\ref{study4pics_fig_soap_signal}). SOAP's interface surfaces only usernames rather than profile pictures, giving users an even narrower basis for distinguishing genuine identities from attacker-supplied alternatives. We examine SOAP's IdP-based binding in detail in Section~\ref{sec:attack}.

Both Vaziripour et al.~\cite{vaziripour2019don} and SOAP~\cite{linker2024soap} share a common property: the service controls which social identities are presented during the ceremony. This is not a limitation of any implementation but a property of the architecture, and the interface's reliance on minimal cues amplifies it, since under service control a carefully constructed fake account is indistinguishable from a genuine one within what the interface reveals. We formalize this in Section~\ref{sec:formal}.

\section{Formal Analysis of Social Authentication Security}
\label{sec:formal}
We characterize ceremony security by a single quantity: whether the interface surfaces any attribute the adversarial service cannot forge, or that the user can check against independent prior knowledge. Let $\mathcal{P}$ denote the set of peers and $\mathcal{X}$ the space of identities, where an \emph{identity} is a set of attribute-value pairs describing a peer's public presence. The genuine-identity mapping is $\mathcal{I}:\mathcal{P}\to\mathcal{X}$, so $\mathcal{I}(P)$ is peer $P$'s genuine identity; $\mathcal{I}_A\in\mathcal{X}$ is an attacker-supplied identity, and $\mathcal{I}_S(P)\in\mathcal{X}$ is whichever identity the directory $S$ delivers for $P$ during the ceremony, equal to $\mathcal{I}(P)$ when $S$ is honest. An \emph{attribute accessor} $a$ is a function $a:\mathcal{X}\to\mathcal{V}_a$ mapping an identity to its value in that attribute's value space $\mathcal{V}_a$; we write $a_{\mathrm{pic}}$, $a_{\mathrm{name}}$, and $a_{\mathrm{user}}$ for the profile picture, display name, and username accessors, and $C$ is the finite set of accessors the ceremony interface surfaces to the user. The user's prior knowledge $K_U$ is a partial map from accessors to genuine values, so $a\in\mathrm{dom}(K_U)$ means the user independently knows $a(\mathcal{I}(P))$. We write $S^*$ for the adversarial instantiation of $S$: one that is compromised, malicious, or coerced, but that cannot compromise any social media provider (the provider stays honest, as in our threat model).

We call a set of attributes $C'\subseteq C$ \emph{forgeable} if $S^*$ can produce a single identity $\mathcal{I}_A\in\mathcal{X}$ with $a(\mathcal{I}_A)=a(\mathcal{I}(P))$ for every $a\in C'$, using only data readable from $P$'s public profile, without compromising any provider. The profile picture, display name, and username are jointly forgeable: all are visible on the same public profile and can be copied verbatim. An attribute is \emph{unforgeable} if matching its genuine value requires a witness $S^*$ cannot produce without detectable misbehavior, such as a client-verifiable key-transparency inclusion proof. Forgeability is thus a property of the set and not of its members: no set containing an unforgeable attribute is forgeable, and a set may fail to be forgeable even when each member is individually reproducible, since one $\mathcal{I}_A$ must satisfy every constraint at once. This split is grounded in the threat model, not in an assumed cost: forgeable means reproducible under an honest social media provider, unforgeable means not. We call an attribute \emph{distinguishing} if it is either unforgeable, or an attribute the user can check against independent prior knowledge $K_U$, and write $\mathcal{D}(C, K_U)$ for the set of distinguishing attributes surfaced by $C$.

\begin{definition}[Social Authentication Ceremony]
\label{def:ceremony}
A ceremony is a tuple $(U, P, S, C, K_U, \mathsf{Ver})$ where $U$ is the verifying user; $P$ the intended peer; $S$ the service directory that delivers the identity $\mathcal{I}_S(P)$; $C$ the set of attribute accessors the interface surfaces (e.g., $a_{\mathrm{pic}}$, $a_{\mathrm{name}}$); $K_U$ the user's prior knowledge of $P$'s genuine identities, obtained independently of $S$ (possibly empty); and $\mathsf{Ver}\big(U, \{a(\mathcal{I}_S(P)) : a \in C\}, K_U\big) \to \{0,1\}$ the accept/reject decision, which depends only on the surfaced attribute values and $K_U$, not on inspection outside the interface.
\end{definition}

\begin{condition}[Identity Binding Integrity]
\label{cond:ibi}
A ceremony is secure against an adversarial service $S^*$ only if (i) $S$ is honest and delivers $\mathcal{I}_S(P)=\mathcal{I}(P)$, or (ii) $\mathcal{D}(C,K_U)$ $\neq\emptyset$: at least one surfaced attribute is unforgeable, or can be checked against the user's prior knowledge $K_U$. If neither holds, $\mathsf{Ver}$'s input is independent of whether the displayed identities are genuine.
\end{condition}

\begin{proposition}
\label{prop:violation}
\attackname\ succeeds with probability equal to the benign acceptance rate whenever the surfaced cue set $C$ is forgeable and the distinguishing set $\mathcal{D}(C,K_U)$ $=\emptyset$. In the deployed interfaces of Vaziripour et al.~\cite{vaziripour2019don} and SOAP~\cite{linker2024soap}, the surfaced cue set is contained in the public-profile accessors, $C \subseteq \{a_{\mathrm{pic}}, a_{\mathrm{name}}, a_{\mathrm{user}}\}$ (profile picture and display name in the former, username in the latter), and these cues are jointly forgeable, and the target scenarios give $K_U=\emptyset$; hence $\mathcal{D}=\emptyset$.
\end{proposition}

\begin{proof}
Suppose $C$ is forgeable and $\mathcal{D}(C,K_U)=\emptyset$. Then $S^*$ produces a single $\mathcal{I}_A\in\mathcal{X}$ with $a(\mathcal{I}_A)=a(\mathcal{I}(P))$ for all $a\in C$. The value set $\{a(\mathcal{I}_A):a\in C\}$ that $\mathsf{Ver}$ receives is then identical to $\{a(\mathcal{I}(P)):a\in C\}$, and, since $\mathcal{D}(C,K_U)=\emptyset$ gives $C\cap\mathrm{dom}(K_U)=\emptyset$, $K_U$ holds no genuine value against which a surfaced attribute can be checked. $\mathsf{Ver}$'s input is therefore the same in both cases, so
\[
\Pr[\mathsf{Ver}(U,\{a(\mathcal{I}_A):a\in C\},K_U)=1]=\Pr[\mathsf{Ver}(U,\{a(\mathcal{I}(P)):a\in C\},K_U)=1],
\]
the benign acceptance rate. The deployed interfaces satisfy both hypotheses, so the equality applies.
\end{proof}

\textbf{Scope of the equality.} Proposition~\ref{prop:violation} equates two acceptance probabilities over $\mathsf{Ver}$'s inputs, not two finite samples. Our measured rates (93.33\% under attack, 96.33\% benign) differ by less than sampling error, which is what it anticipates: the ceremony contributes no signal, though trial-level noise remains.

\textbf{Reconciliation with Prior Designs.} Vaziripour et al.\ and SOAP were not careless: their designs are secure if $\mathcal{D}\neq\emptyset$, which their designs implicitly assume, on the premise that a genuine account's accumulated history is available to the verifier and that platform-side detection filters accounts that lack it. This premise fails for two reasons. The interface surfaces only a thin cue set $C$, so that history is not in $C$ and never reaches $\mathsf{Ver}$; and because the adversarial service is the channel delivering identities, it bypasses platform-side detection entirely. Condition~\ref{cond:ibi} makes this implicit premise explicit and shows it is unsatisfied, rather than attributing the weakness to a platform detection failure. Security is therefore restored only by making $\mathcal{D}\neq\emptyset$, through a surfaced unforgeable attribute such as a client-verifiable transparency proof, or a guaranteed out-of-band $K_U$; adding more forgeable cues does not.

Section~\ref{sec:attack} maps $\mathcal{D}$ across existing systems. Section~\ref{sec:study} evaluates Proposition~\ref{prop:violation} empirically. Section~\ref{sec:discussion} derives mitigations from Condition~\ref{cond:ibi}.

\section{Our \attackname\ Attack Framework}
\label{sec:attack}
\attackname\ operationalizes the violation of Condition~\ref{cond:ibi} from Section~\ref{sec:formal}. We describe the attack mechanism, its implications, and the threat model under which it operates, targeting end-to-end encrypted applications like Signal~\cite{signal} and WhatsApp~\cite{whatsapp} under the threat model in Subsection~\ref{study4_threat_model}.

\subsection{Description of the Attack}
The adversary is a malicious, coerced, or compromised service provider that mounts a MitM attack by distributing fake public keys through fake social media accounts during the ceremony. In the social authentication of Vaziripour et al.~\cite{vaziripour2019don}, users register the social media accounts they control, the application publishes their public key on those accounts, and during the ceremony it compares a peer's key against the keys advertised on the peer's linked accounts; a match is surfaced for the user to confirm.

Our simulation did not deploy production-grade implementations of Vaziripour et al.~\cite{vaziripour2019don} or SOAP, but mirrored the core protocol operations common to both: key exchange, identity verification through social media profiles, and server-side key management. The simulated keyserver dynamically distributed counterfeit keys and bound them to fake accounts, enabling a MitM attack against the identity-binding assumptions both designs share.

\begin{figure}[!t]
    \centering
    \includegraphics[width=0.75\linewidth,height=5cm]{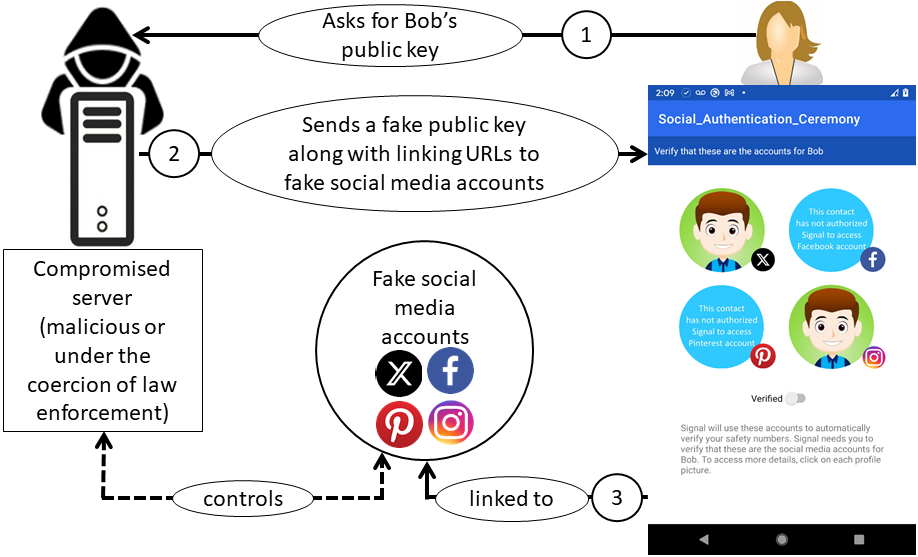}
    \caption{An overview of our proposed {\attackname} Attack Framework.}
    \label{study4pics_fig1}
\end{figure}

The malicious server substitutes the legitimate key with a fake one and links the peer's profiles to fake accounts advertising that same fake key, so the server-promoted and account-advertised keys match; the attributes the attacker must reproduce ($a_{\mathrm{pic}}$, $a_{\mathrm{name}}$, and $a_{\mathrm{user}}$ where surfaced) are jointly forgeable, since each is copied directly from the peer's public profile. Figure~\ref{study4pics_fig1} shows the framework, with steps below:

\begin{enumerate}
	\item Alice initiates communication with Bob through an E2EE application like Signal~\cite{signal}, requesting Bob's public key from the server.
	\item The compromised or malicious server returns a fake key for Bob with URLs to fabricated social media accounts, falsely claiming the key belongs to Bob.
	\item During the ceremony, Alice is shown profiles supposedly belonging to Bob but in fact controlled by the server. Inspecting them within the interface, Alice unwittingly accepts the fraudulent accounts.
\end{enumerate}

This mechanism violates both clauses of Condition~\ref{cond:ibi}. Clause~(i) fails because the server deliberately delivers $\mathcal{I}_A \neq \mathcal{I}(P)$. Clause~(ii) fails because the surfaced cue set ($a_{\mathrm{pic}}$, $a_{\mathrm{name}}$) is jointly forgeable and the deployment scenarios give $K_U = \emptyset$, so $\mathcal{D} = \emptyset$ and $\mathsf{Ver}$ receives identically distributed inputs under $\mathcal{I}_A$ and $\mathcal{I}(P)$. In our study, where the accounts available for inspection were researcher-built, click-through did not improve detection (Section~\ref{sec:results}). We do not claim inspection cannot help in general: surfacing an unforgeable attribute, such as a genuine account's verifiable history, would place a distinguishing attribute within reach, which our deployed-interface model did not provide.

We evaluate the profile-based ceremony of Vaziripour et al.~\cite{vaziripour2019don} empirically and treat SOAP~\cite{linker2024soap} analytically rather than reimplementing its OpenID Connect flow. SOAP binds identities to IdP-issued tokens, so its security rests on the IdP, not on the service selecting accounts. This attestation protects the token-to-identifier binding but certifies only that some party controls an identifier, not that the identifier belongs to the intended peer. The verifier must still judge whether the attested username shown is the peer's. A username is forgeable, so a service that controls which IdP-attested identity is displayed can present an attacker's genuinely attested account as the peer's; when $K_U = \emptyset$, the cue set contains no distinguishing attribute and Condition~\ref{cond:ibi} fails at the display layer. Confirming this against SOAP's deployed flow requires reimplementing its IdP-mediated verification, which we leave to future work.

\subsection{Implications of \attackname}
\attackname\ differs from conventional phishing because the manipulation occurs entirely inside the trusted application interface. Users are never redirected outside the application and see only the identities the service chooses to display. The vulnerability is architectural: it stems from service-mediated identity presentation, not from user inattentiveness or platform-side detection failures. Improving a platform's ability to detect fake accounts does not mitigate \attackname, because the adversarial service introduces attacker-controlled identities through the ceremony itself, bypassing platform policies.

\textbf{Threats to Social Media Integration with E2EE Authentication Ceremonies.} Vaziripour et al.~\cite{vaziripour2019don} and Linker and Basin~\cite{linker2024soap} both place identity selection under service control. As discussed in Section~\ref{sec:formal}, this reflects an implicit premise that the surfaced cues carry unforgeable signal, unsatisfied under service control, rather than a platform-side detection failure. The interface amplifies the weakness by surfacing only minimal cues with no path for out-of-app validation, more so in SOAP~\cite{linker2024soap}, where verification relies on usernames alone. Fake accounts for fraudulent activity are common~\cite{wani2017sneak,kumar2018false,apte2019frauds}, yet their impact on E2EE key authentication is unexamined. Mink et al.~\cite{mink2022deepphish} found users accepted AI-generated fake LinkedIn~\cite{linkedin} profiles at 79\% to 85\%, supporting \attackname's premise that users trust visually plausible profiles even on open platforms richer in cues than a constrained ceremony interface.

\textbf{Threats to Vouching and Trusted Friend Systems.} Brainard et al.~\cite{brainard2006fourth} proposed a vouching authenticator in which a helper issues a temporary \textit{vouchcode} for emergency access, with security depending on the helper recognizing the peer. \attackname\ breaks this through the same mechanism: fake accounts mimicking the peer can deceive the helper into releasing the vouchcode without credential compromise, showing the attack surface generalizes to any system relying on social identity recognition as a trust anchor.

\subsection{Threat Model and Attacker Assumptions}
\label{study4_threat_model}

\begin{table*}[t]
\centering
\caption{Vulnerability of Social Authentication Systems to \attackname\ Under Adversarial Service Control}
\footnotesize
\setlength{\tabcolsep}{4pt}
\renewcommand{\arraystretch}{1.2}
\begin{tabularx}{\textwidth}{|X|X|c|c|X|}
\hline
\textbf{System} & \textbf{Identity Presentation Controlled By} & \textbf{Key Transp.?} & \textbf{Prior Knowl.?} & \textbf{Vulnerable to \attackname?} \\
\hline
Vaziripour et al.~\cite{vaziripour2019don} & Service directory & No & No & Yes \\
\hline
SOAP~\cite{linker2024soap} & Service + IdP & No & No & Analytical: IdP attests identifier control, not peer binding (not evaluated) \\
\hline
Keybase~\cite{keybase} & User-controlled posts & No & Partial & Partial (directory binding layer) \\
\hline
WhatsApp Key Transparency~\cite{whatsapp} & Client-verifiable log & Yes & No & Mitigated \\
\hline
\end{tabularx}
\label{tab:threatmodelcases}
\end{table*}

Social authentication burdens users with recognizing a peer's correct social identities even when they hold no reliable prior knowledge of them, and the interface offers no metadata, cross-platform context, or out-of-app validation to support that judgment. This is the structural problem of Section~\ref{sec:formal}: under service control the verification task becomes ill-defined. Table~\ref{tab:threatmodelcases} maps the vulnerability across systems, showing that any design placing identity presentation under service control is susceptible, while client-verifiable key transparency substantially mitigates it. Our threat model assumes adversarial control over the service-operated directory delivering peer identities and keys, consistent with prior work on social authentication and directory equivocation~\cite{herzberg2016can,yadav2022automatic}. It covers scenarios increasingly common today: new professional contacts, cross-platform authentication without a prior relationship, and pseudonymous communication, where users cannot rely on pre-existing relationships and must depend on whatever cues the service surfaces.

\textbf{Adversary Capabilities.} The adversary has three capabilities: (1) server-side control over key distribution and the presented identity set; (2) the ability to create short-lived but valid profiles that persist long enough for the ceremony without long-term reputation signals; and (3) the ability to populate them with public information about the peer. The attacker does not control the client, and no credentials at any social media provider are compromised; the provider is assumed honest throughout. When Alice contacts Bob, the attacker suppresses Bob's real key and presents a fake key bound to attacker-controlled profiles; if Alice is deceived, the application confirms the match. The deception is symmetric when Bob initiates. If both accept, the ceremony concludes under attacker control and the adversary can intercept, read, and modify all session traffic in real time. Any design in which $S$ controls identity selection violates clause~(i); any design whose surfaced cues are all forgeable and is deployed without prior relationships violates clause~(ii). Both Vaziripour et al.~\cite{vaziripour2019don} and SOAP~\cite{linker2024soap} fall into both categories. Section~\ref{sec:study} describes the user study evaluating whether participants detect \attackname\ under realistic interface constraints.

\begin{figure}[!t]
	\centering
	\includegraphics[width=0.3\textwidth, height=6cm]{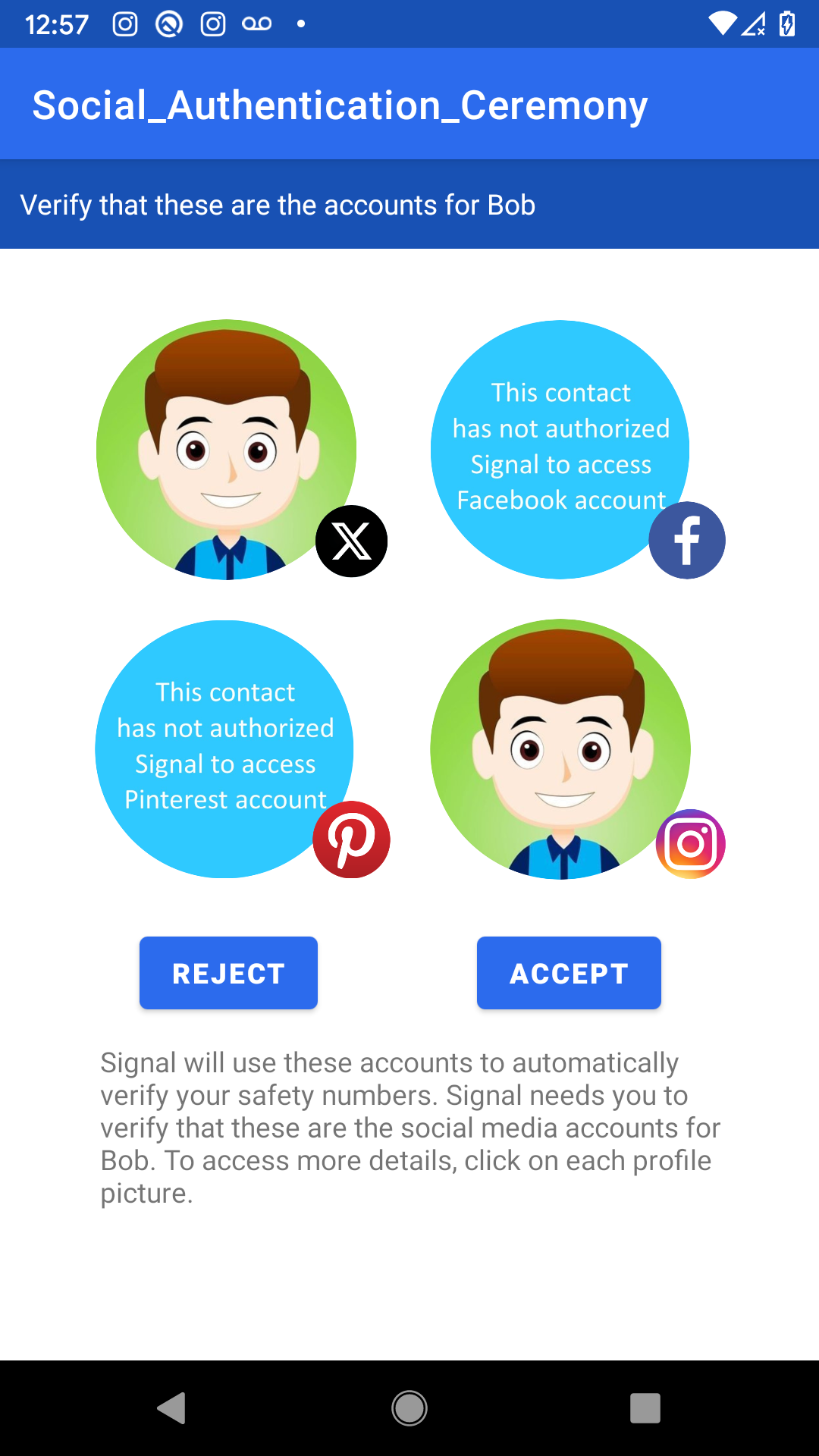}
	\caption{The user interface of the phone application used in the user study.} \label{study4pics_fig4}
\end{figure}

\section{Assessing \attackname: User Study Design}
\label{sec:study}
To evaluate users' responses to \attackname, we conducted a controlled study instantiating the attack in a simulated authentication ceremony based on \cite{vaziripour2019don}. The study tests feasibility: whether a low-effort attacker can construct accounts that survive user inspection within the constrained ceremony interface, not population-level usability.

\subsection{Implementation}
We redesigned the ceremony following \cite{vaziripour2019don}, implementing a dedicated Android application and a simulated malicious server that behaves as the compromised keyserver assumed in both SOAP and the social-media-based mechanism, dynamically generating and distributing fake keys. The simulation replicated the verification processes common to both, including authenticating keys against presented profiles; while not a live deployment, it preserved fidelity to their security and interaction models.

\textbf{Phone Application and Simulated Malicious Signal Server.} Upon initiating the ceremony, the application presents social media profiles associated with the recipient's accounts, and users accept or reject them. On acceptance, the application compares the recipient's public key with the keys from the linked accounts; a match confirms the identity. Our aim was to assess whether users could detect identity substitution under the interface's constraints, not to measure distinguishability between fully established and fabricated accounts. We included two buttons (\textit{accept}, \textit{reject}) to streamline decisions, as our focus was feasibility rather than usability. Figure~\ref{study4pics_fig4} shows the application.

Because posting keys directly to social media requires platform approval, we followed \cite{vaziripour2019don} and stored keys on a separate server mapping each account identifier to its key, including fake keys linked to fabricated accounts, mirroring the malicious server's mapping.

\textbf{Creating Fake Social Media Accounts.} All accounts in the study, both the reference accounts representing the peer and the attacker accounts, were researcher-created; no participant personal accounts or pre-existing established accounts were used. The fabricated accounts adhered to the following criteria: realistic names and profile pictures; content relevant to the target user; following users pertinent to the target; using the target's language; and posting a fake public key retrievable for comparison against the counterfeit key promoted by the malicious server.

\begin{figure*}[h]
	\centering
	\includegraphics[width=0.80\linewidth, height=6cm]{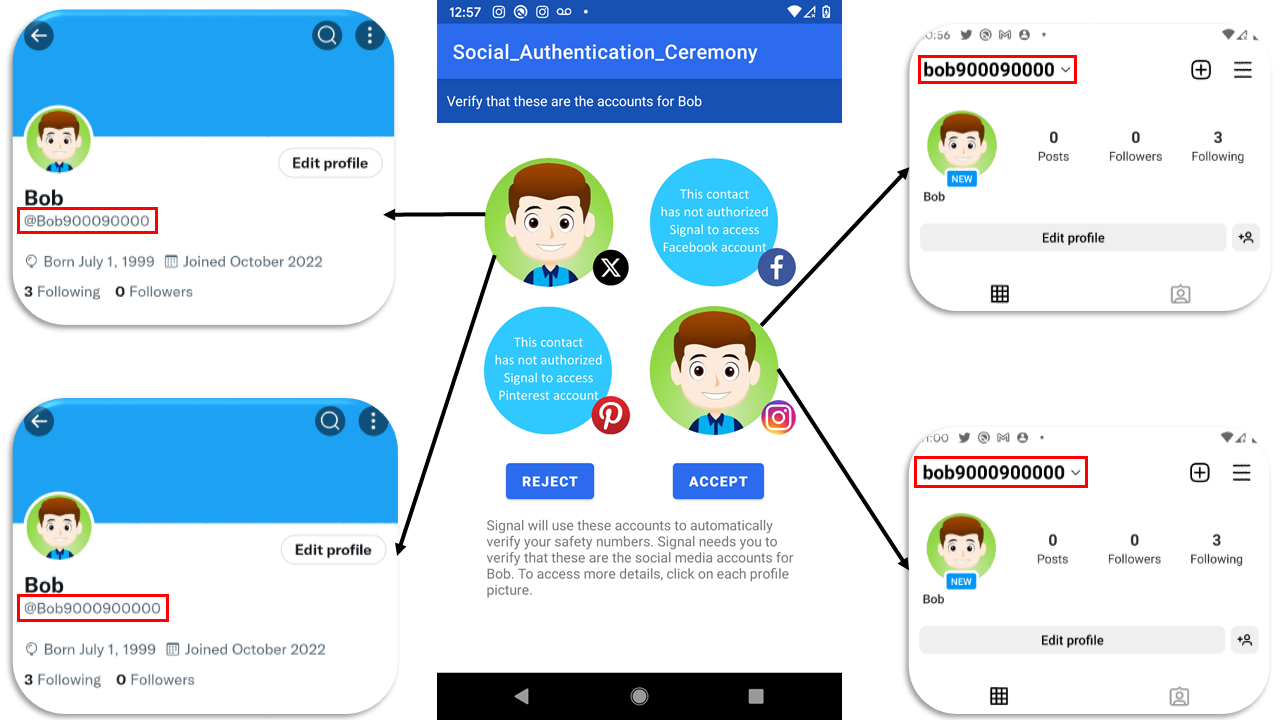}
	\caption{Snapshots of linking original and fake social media accounts to a specific user during the authentication ceremony} 
    \label{study4pics_fig3}
\end{figure*}

In the \attackname\ scenario (Figure~\ref{study4pics_fig3}), we created two accounts per user on X~\cite{x} or Instagram~\cite{instagram}. The first was a researcher-built reference account standing in for the peer's genuine profile, posting the real key; the second, with fabricated details targeting the victim, posted the fake key on the same platform. During the ceremony we linked each user's profile to a corresponding fake account on every occasion. The fabricated profiles closely resembled the reference, and the fake keys matched the counterfeit key promoted by the malicious server, completing the substitution.

\subsection{Authentication Ceremony Sessions} We established three session types: \textbf{Benign}, in which researcher-built reference profiles representing the peer are displayed and linked to their accounts, serving as a manipulation check on task comprehension rather than a genuine-versus-fake discrimination test, since the reference accounts were themselves researcher-created (Section~\ref{sec:discussion}). The reference and \attackname\ profiles are intentionally matched on the attributes surfaced by the ceremony, placing $\mathsf{Ver}$ in the empty-distinguishing-set regime of Proposition~\ref{prop:violation}; \attackname, in which counterfeit profiles matching the reference on picture and name are linked to fabricated accounts; and \textbf{Baseline Attack}, in which unrelated profiles are linked to unrelated accounts.

\subsection{Study Protocol}
\textbf{Ethical Considerations.} The study received IRB approval; participation was voluntary with opt-out at any point, and no identifiable information was stored. We recruited 30 participants across age groups, all familiar with social media, each given an Android smartphone. A researcher facilitated, explaining the ceremony and the use of social media for posting and retrieving keys.

\textbf{Orientation.} The facilitator explained the ceremony, demonstrated posting and retrieving keys, and showed how to open each profile on X~\cite{x} and Instagram~\cite{instagram}. Because personal accounts were not used, the facilitator created all reference and counterfeit accounts. To avoid bias, only the mechanics were explained, with no examples of fake accounts.

\textbf{Pre-test.} Participants completed a demographic questionnaire on age, gender, and social media familiarity (Appendix~\ref{appendixA}).

\textbf{Familiarization.} Participants practiced by verifying whether displayed profiles belonged to the intended recipient, Bob, visiting each linked account across six benign trials, with randomized request timing and varied usernames. No attack examples were provided.

\textbf{Data Collection.} Participants completed 40 trials mixing benign and attack scenarios: twenty displayed Bob's reference picture and name, twenty displayed unrelated identities, and among the Bob-like trials half were \attackname\ trials reusing the reference picture and name. Because the interface surfaces only a picture and display name, both reproduced by the counterfeit accounts, no interface-visible attribute distinguished an attacker account from the reference; with $K_U = \emptyset$, this places $\mathsf{Ver}$ in the empty-distinguishing-set regime of Proposition~\ref{prop:violation} and lets us test whether the attack succeeds in practice.

\textbf{Post-test.} The post-test questionnaire collected feedback on suspected behavior, reasons for decisions, and how often participants opened profile links (Appendix~\ref{appendixB}).

\subsection{Study Metrics}
We use two primary metrics. \emph{Benign Success Rate} (BSR) is the rate at which users accept the reference profiles, functioning as a manipulation check that participants engaged with the task rather than accepting indiscriminately, not as a genuine-versus-fake discrimination measure. \emph{Attack Success Rate} (ASR) is the rate at which users accept counterfeit profiles linked to fabricated accounts of the peer; a high ASR indicates that low-effort counterfeit accounts pass inspection, and the gap between ASR and the baseline isolates this effect.
\begin{equation}\label{Eq:BSR}
BSR = \frac{\#accepted\_benign\_sessions}{\#benign\_sessions}, \quad
ASR = \frac{\#accepted\_attack\_sessions}{\#fake\_account\_attack\_sessions}
\end{equation}

As a baseline, we measure how often users accept entirely unrelated profiles linked to unrelated accounts, expecting this to approach 0\%:
\begin{equation}\label{Eq:Baseline}
	Baseline = \frac{\#accepted\_baseline\_attack\_sessions}{\#different\_account\_attack\_sessions}
\end{equation}

\begin{table}[t]
\centering
\caption{Summary of User Study Results}
\label{study4tabs_tab1}
\footnotesize
\setlength{\tabcolsep}{5pt}
\renewcommand{\arraystretch}{1.15}
\begin{tabular}{lccccccc}
\toprule
\multirow{3}{*}{\textbf{Session}} & \multicolumn{2}{c}{\textbf{Response}} & \multirow{3}{*}{\textbf{BSR}} & \multirow{3}{*}{\textbf{Baseline}} & \multirow{3}{*}{\textbf{ASR}} & \multicolumn{2}{c}{\textbf{Response Time (s)}} \\
\cmidrule(lr){2-3} \cmidrule(lr){7-8}
& \textbf{Accept} & \textbf{Reject} & & & & \multirow{2}{*}{\textbf{Mean}} & \multirow{2}{*}{\textbf{SD}} \\
& \textbf{(Verify)} & \textbf{(Disprove)} & & & & & \\
\midrule
Benign          & 96.33\% & 3.67\%  & \textbf{96.33\%} & --       & --       & 30.90 & 11.45 \\
Baseline        & 2.00\%  & 98.00\% & --       & \textbf{2.00\%}  & --       & 12.93 & 6.12  \\
\attackname     & 93.33\% & 6.67\%  & --       & --       & \textbf{93.33\%} & 30.82 & 11.46 \\
\bottomrule
\end{tabular}
\end{table}

\section{Results and Analysis}
\label{sec:results}
\textbf{Demographic Information.}
Of the 30 participants, 53.4\% were aged 21--30, 40\% aged 31--40, and 6.6\% above 40; 63.3\% were male and 36.7\% female. Academic backgrounds were diverse, with 70\% holding or pursuing a graduate degree, 83.4\% reporting proficient security skills, and an average self-reported social media familiarity of 4.47 out of 5. This composition is analytically significant: a technically skilled, security-aware cohort with high social media fluency is the group \emph{least likely} to fall for an identity substitution attack, so the observed ASR is a conservative lower bound, with broader populations likely at least as susceptible.

\textbf{Benign and Baseline Sessions.}
The benign success rate was 96.33\% (95\% CI: [81.0\%, 99.9\%]), confirming participants understood the task and engaged with the ceremony rather than accepting indiscriminately. The baseline rejection rate was 98\%, with users accepting only 2\% of clearly mismatched profiles (95\% CI: [0.05\%, 10.7\%]), confirming they could distinguish obviously different individuals. Table~\ref{study4tabs_tab1} summarizes all sessions.

\textbf{Attack Sessions.}
The attack success rate was 93.33\% (28 of 30 participants; 95\% CI: [77.9\%, 99.2\%]). Similarity between benign and \attackname\ acceptance is expected from the matched construction; the informative empirical contrast is therefore with the Baseline Attack: participants rejected obviously mismatched profiles (2\% acceptance) but accepted carefully constructed substitutions satisfying $a(\mathcal{I}_A)=a(\mathcal{I}(P))$ on every interface-visible attribute (93.33\% acceptance), a large and significant separation (WSRT $p<0.001$, $r=0.54$).

This is exactly the outcome Proposition~\ref{prop:violation} predicts for the empty-distinguishing-set regime instantiated by the study: when $K_U=\emptyset$ and the surfaced cues are all forgeable, the interface offers no attribute by which $\mathcal{I}_A$ can be told from $\mathcal{I}(P)$.

Post-study surveys recorded zero reported suspicion during attack sessions; all reported suspicion referred to baseline cases, where rejection was attributed to mismatched profile pictures (80\% also cited divergent names, 30\% differing identifiers).

\textbf{Behavioral Engagement.}
We recorded whether participants opened the linked accounts during the ceremony: 13.3\% never did, 20\% did sporadically, 40\% often, and 26.7\% on every trial. The majority relied on the in-app profile picture, which partly explains the high ASR, since matching pictures leave in-app verification no basis for detection. More importantly, the attack succeeded at high rates even among those who clicked through on every trial; the aggregate 93.33\% holds across the full pool including the most thorough inspectors. This is corroborated by response times: benign and attack sessions took near-identical time (30.90\,s, SD 11.45; and 30.82\,s, SD 11.46; a 0.08\,s difference, WSRT $p=0.1047$), both substantially longer than the baseline session (12.93\,s, SD 6.12; $p=0.0157$ vs.\ attack). Participants thus deliberated over attack trials as carefully as benign ones rather than accepting hastily, yet still failed to detect substitution. The failure is architectural, not behavioral: when the service controls identity presentation and the interface surfaces no unforgeable cue, no amount of engagement supplies a basis for detection. We discuss mitigation in Section~\ref{sec:discussion}.
\section{Discussion}
\label{sec:discussion}
\textbf{User Study Limitations.}
Our study evaluates the feasibility of \attackname, not population-level usability. Two limitations follow from its design. First, all accounts were researcher-created rather than participants' personal or pre-existing established accounts; the benign condition therefore functions as a manipulation check on task comprehension, not as a genuine-versus-fake discrimination test. Because the ceremony interface surfaces only a profile picture and display name, an established account's richer history never reaches the verifier in the in-app path, so the in-app result is unaffected by this choice; the limitation bears on the click-through path, where participants inspected researcher-built rather than long-lived accounts, and the attack may be somewhat easier there than against a peer with a rich, hard-to-forge presence. We note, however, that in the target scenarios (new contacts, pseudonymous communication) the peer's real account is often itself sparse, narrowing this gap. Thus, our study does not claim that counterfeit accounts are indistinguishable from long-established genuine accounts under unrestricted external inspection; it evaluates whether substitution remains effective when the authentication ceremony itself exposes no distinguishing cue. Second, participants completed forty trials, raising possible repetition effects; randomized timing, varied profile content, and mixed benign and attack trials mitigate simple habituation.

The cohort is conservative: a technically skilled, security-aware group with high social media fluency is the population least likely to fall for this attack, so its failure rate is a lower bound for broader, less experienced populations. The consistent failure across benign, baseline, and attack trials points to a design constraint inherent in surfacing only minimal identity cues at verification time, not to any feature of this pool.

\textbf{Potential Defenses.}
By Condition~\ref{cond:ibi}, a mitigation helps only if it makes the distinguishing set non-empty, which interface tweaks that add further forgeable cues cannot do. Surfacing richer but still copyable attributes (profile name, identifier, metadata), as in Meta's key-transparency rollout for Messenger~\cite{meta-messenger-key-transparency-2025}, and friction nudges that slow shallow click-through~\cite{nist-sp800-63-4} both help at the margin but are insufficient: among participants who inspected every linked account, the attack still succeeded, because a well-built fake presents the same signals as a genuine one under any inspection the interface affords. The only defenses that restore security constrain the service itself. Binding social-media attestations to an auditable key-transparency log lets clients detect silent substitution~\cite{malvai2023parakeet}; user-controlled or decentralized identifiers resist silent remapping~\cite{w3c-did-core-2022}; and an out-of-band confirmation channel supplies an independent reference~\cite{openid-identity-assurance-2024}. Anomaly detection on bursts of ephemeral accounts~\cite{shukla2025fraudulent} offers only complementary protection, since it cannot prevent substitution when the service controlling which accounts are surfaced is itself adversarial. The path to secure social authentication runs through architectural reform: only removing service-controlled identity selection from the ceremony's trust boundary resolves the vulnerability of Condition~\ref{cond:ibi}.

\section{Other Related Work}
\label{relatedworks}
\textbf{Authentication Ceremony Security and Usability.}
A substantial body of work examines why ceremonies fail. Shirvanian and Saxena showed short authentication strings in crypto-phone ceremonies are vulnerable to voice-imitation MitM, reaching $\sim$50\% indistinguishability via voice morphing~\cite{shirvanian2014wiretapping}, and documented 30--40\% adversarial success on VoIP checksum comparison from hasty verification~\cite{shirvanian2015security}. Shirvanian et al.~\cite{shirvanian2017pitfalls} found systematic failures in remote fingerprint verification under time pressure. On usability, Dechand et al.~\cite{dechand2016empirical} showed fingerprint encoding choices drive verification accuracy, with some representations exceeding 50\% error, and Tan et al.~\cite{tan2017can} found visual encodings only modestly better than textual. Unlike these channels, \attackname\ targets the social identity presentation layer and succeeds regardless of how carefully users engage.

\textbf{Automated Fake Key Detection.}
Yadav et al.~\cite{yadav2022automatic} detect fake-key attacks using client-verifiable consistency logs, making malicious-server key substitution detectable when clients maintain auditable key-to-account mappings. \attackname\ evades this entirely: the substituted keys are self-consistent within the adversarial directory, introducing no cryptographic inconsistency, because the vulnerability lives at the identity presentation layer that transparency logs do not constrain.

\textbf{Fake Accounts and Social Identity Deception.}
\attackname's feasibility depends on visually plausible fake accounts that survive inspection, which prior work confirms is achievable at scale: short-lived, purpose-built accounts are operationally common~\cite{wani2017sneak,kumar2018false,apte2019frauds}, and Shukla et al.~\cite{shukla2025fraudulent} show inauthentic-account detection relies on longitudinal signals unavailable for ephemeral profiles. \attackname\ exploits this gap directly, since its fake accounts need only persist for the duration of the ceremony.
\section{Conclusion}
Social authentication made E2EE key verification usable by moving the trust anchor from a cryptographic primitive with a deterministic answer to a service-mediated identity judgment with no reliable basis for resolution: the service that authentication exists to constrain decides what users see. \attackname\ shows this is directly exploitable, with no credential compromise, lure, or implementation flaw, defeating even users who inspect thoroughly. The remedy is therefore not interface polish but architectures that remove the service from the identity-delivery trust boundary, through client-verifiable transparency logs, user-controlled attestations, or decentralized identifiers. Until then, social authentication offers only a false sense of security: users believe they are verifying their peer while in fact trusting the service to show them the right one.

\section*{Acknowledgements}
We thank the reviewers for their invaluable comments and
constructive feedback, which substantially improved this paper.
\bibliographystyle{splncs04}
\bibliography{references}

\appendix
\section{Pre-Test Questionnaire Questions}
\label{appendixA}
\noindent\textbf{Q1.} Gender? (1)~Male \quad (2)~Female

\noindent\textbf{Q2.} Age group? (1)~$\leq$20 \quad (2)~21--30 \quad (3)~31--40 \quad (4)~41--50 \quad (5)~$\geq$51

\noindent\textbf{Q3.} Highest education completed? (1)~Bachelor's \quad (2)~Master's \quad (3)~Doctoral \quad (4)~None of the above

\noindent\textbf{Q4.} Familiarity with social media networks, 1 (least) to 5 (most): (1)~\, (2)~\, (3)~\, (4)~\, (5)

\noindent\textbf{Q5.} General security skills, 1 (least) to 5 (most): (1)~\, (2)~\, (3)~\, (4)~\, (5)

\section{Post-Test Questionnaire Questions}
\label{appendixB}
\noindent\textbf{Q1.} Did you notice any suspicious behavior in the application? (1)~Yes \quad (2)~No \quad (3)~Don't remember

\noindent\textbf{Q2.} Possible reason for the suspicious behavior (if any)? (1)~Malicious UI activity \quad (2)~System-generated fake profiles \quad (3)~No idea \quad (4)~Noticed none

\noindent\textbf{Q3.} Did you reject any social media profiles? (1)~Yes \quad (2)~No \quad (3)~Don't remember

\noindent\textbf{Q4.} If you rejected a profile, reasons (select all; ``N/A'' if you approved all): (1)~Different account name \quad (2)~Different profile picture \quad (3)~Different unique identifier \quad (4)~N/A

\noindent\textbf{Q5.} How often did you check the profile URL during the study? (1)~Never \quad (2)~$<$40\% \quad (3)~40--60\% \quad (4)~60--90\% \quad (5)~Always

\end{document}